\documentclass [12pt]{article}
\usepackage{graphicx}
\usepackage{bm}

\begin{document}

%\title[]{}
\noindent
{\bf How is entropy production rate related to chemical reaction rate?}

\vskip .1in

Kinshuk Banerjee and Kamal Bhattacharyya\footnote{Corresponding 
author; e-mail: pchemkb@yahoo.com}

\vskip .05in
{\it Department of Chemistry, University of Calcutta, 
92 A.P.C. Road,

Kolkata 700 009, India.}

%\date{\today}

\begin{abstract}

The entropy production rate is a key quantity 
in irreversible thermodynamics. 
In this work, we concentrate on the realization of 
entropy production rate in chemical reaction systems in terms 
of the experimentally measurable reaction rate. 
Both triangular and linear 
networks have been studied. They attain either thermodynamic equilibrium 
or a non-equilibrium steady state, under suitable external 
constraints. We have shown that the 
entropy production rate is proportional to the square 
of the reaction velocity only around equilibrium 
and not any arbitrary non-equilibrium steady state. 
This feature can act as a guide in revealing the 
nature of a steady state, very much like 
the minimum entropy production principle. 
A discussion on this point has also been presented. 

\end{abstract}

keywords: Entropy production rate, reaction rate, non-equilibrium 

steady state

%\maketitle

\section{Introduction}

Twentieth century witnessed a paradigm shift in the field of 
thermodynamics. The focus of the scientific community 
gradually changed from equilibrium 
thermodynamics of the previous era to the thermodynamics of 
irreversible processes \cite{ons1,ons2} 
and of steady states \cite{denb}. 
Starting with the pioneering works of Onsager in the form 
of reciprocal relations in coupled irreversible processes 
\cite{ons1,ons2}, 
research in non-equilibrium thermodynamics 
expanded rapidly \cite{groot}. 
The power of the subject to capture real, natural processes 
ensured its multidisciplinary nature \cite{groot1} and its 
applicabilty to chemistry, physics, biology as well as to various 
technological aspects \cite{katch}. 
Over the years, the theoretical tools and understanding improved 
and expanded in various directions \cite{prigogn1}. 
The linear laws of Onsager, applicable to states near 
thermodynamic equilibrium (TE), 
were generalized by Prigogine and coworkers giving rise 
to non-linear, irreversible thermodynamics for 
states far removed from the TE \cite{prigogn2,prigogn3}. 
%Their theory showed the emergence of 
%`dissipative structures', in open systems arising 
%out of disorder\cite{prigogn3}. 
In the last two decades, the theory has evolved into 
the thermodynamics of small systems \cite{jarz2011}. 
The fundamental role 
of fluctuations in governing the properties of these 
systems has been revealed and the link between 
microscopic reversibilty and macroscopic irreversibility 
is established in terms of the fluctuation theorems 
\cite{sevk,seifert}. 
Chemical reaction systems have also been treated extensively 
under this field, 
going beyond the realm of TE \cite{gasprd,gasprd1,hqian}. 
Various analytical and numerical methodologies 
have emerged to study the non-equilibrium thermodynamics 
\cite{Niven} 
of reactions occuring in bulk as well as at the 
level of few molecules \cite{Hge}, along with their kinetics
\cite{Gilsp,Xiao}.

In all these developments, entropy plays the 
part of the most basic and interesting thermodynamic 
quantity \cite{Schnak}. 
A quintessential thermodynamic feature of a system 
out-of-equilibrium or an open system 
is the emergence of a non-equilibrium steady state (NESS) 
\cite{Min,Zhang,Jiang,KBan}, 
with the state of equilibrium being a special case. 
Whether NESS prevails in a certain situation 
is characterized by 
non-vanishing entropy production rate (EPR) \cite{Galavotti}, 
measuring the dissipation 
associated with the process \cite{konde}. 
Still, the measure is a theoretical one and, for 
complex systems, the 
connection of EPR with the immediately observable quantities 
may not be apparent.  
Therefore, it will be helpful to get an idea about 
how the EPR of an irreversible process 
is connected with some ready experimentally observable 
quantity. 
In this context, here we study the EPR in relation to 
chemical reaction rates. 
We consider a triangular as well as a linear reaction network. 
The choices are dictated primarily by simplicity, yet 
exhibiting nontrivial features that 
permit the emergence of NESS \cite{Hqian1,Hqian2}. 
We particularly focus on the possibility of realization of 
EPR in these reaction networks in terms of the 
experimentally measurable 
velocity of the reaction both near the TE as well as NESS. 
It is important to note that the reaction velocity $v(t)$ 
is {\it not generally 
equal to the reaction flux}, conventionally used in the 
definition of EPR \cite{groot,nicol,espo}. This is 
particularly true for cyclic reaction networks, 
justifying its choice as a case study. 
To be specific, here we show that, with $P,Q,R$ as constants,

(i) EPR = $P+Qv(t)+Rv^2(t)$, around a NESS,

(ii) EPR = $Rv^2(t)$, around the TE.

Thus the proportionality of the EPR with square of the reaction rate 
becomes a hallmark of the TE. 
This criterion may well be used to distinguish a NESS from 
the TE. In this issue, mention may be made of 
the work of Ross {\it et al.} \cite{ross1} that established 
such a distinction in terms of the minimum entropy production 
principle (MEPP) \cite{groot1,prigogn2}.
Therefore, we briefly comment on the connection of our findings 
with the MEPP.

\section{Entropy production rate and reaction rate around TE} 

Let us study the EPR of 
a triangular reaction network and its linear counterpart 
around a TE. The relation between the EPR 
and reaction rate is derived in each of the cases near TE. 
Here we consider the cyclic system first, because all the 
results of the simpler linear network would 
follow as a special case of the former.

\subsection{ABC cyclic network}

\begin{figure}[tbh]
\centering
\rotatebox{270}{
\includegraphics[width=6cm,keepaspectratio]{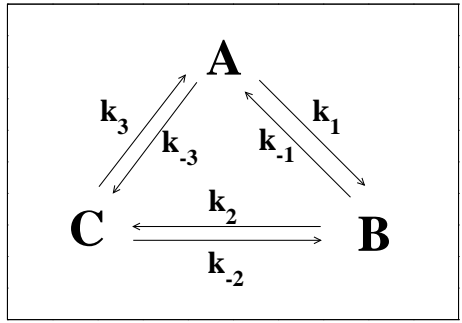}}%{ABC-scheme.ps}}
\caption{Schematic diagram of the ABC cyclic reaction network indicating 
the forward and backward rate constants of each reaction.}
\label{fig1}
\end{figure}

The kinetic scheme of the ABC cyclic network is shown 
in Fig.\ref{fig1}. 
%%%%%%%%%%%%%%%%%%%%%%%%%%%%%%%%%%%%%%%%%%%%%%%%%%%%%%%%%%%%%%%%%%%%%%
%%%%%%%%%%%%%%%%%%%%%%%%%%%%%%%%%%%%%%%%%%%%%%%%%%%%%%%%%%%%%%%%%%%%%%%%%
The kinetic equations of the reaction system are 
written as 
\begin{equation}
\dot{a}=-(k_1+k_{-3})a(t)+k_{-1}b(t)+k_3c(t),
\label{abc1}
\end{equation}
\begin{equation}
\dot{b}=k_1a(t)-(k_{-1}+k_2)b(t)+k_{-2}c(t),
\label{abc2}
\end{equation}
\begin{equation}
\dot{c}=k_{-3}a(t)+k_2b(t)-(k_{-2}+k_3)c(t),
\label{abc3}
\end{equation}
with $a(t), b(t), c(t)$ being the 
concentrations of species A, B, C, respectively, at time $t$. 
At the steady state, $\dot{a}=\dot{b}=\dot{c}=0.$ 
Then one obtains the steady-state solutions as
\begin{equation}
a^s=(k_2k_3+k_{-1}k_3+k_{-1}k_{-2})/N_1=\alpha/N_1,
\label{as}
\end{equation}
\begin{equation}
b^s=(k_1k_3+k_{1}k_{-2}+k_{-3}k_{-2})/N_1=\beta/N_1,
\label{bs}
\end{equation}
\begin{equation}
c^s=(k_1k_2+k_{2}k_{-3}+k_{-1}k_{-3})/N_1=\gamma/N_1,
\label{cs}
\end{equation}
with $N_1=\alpha+\beta+\gamma.$

The EPR $\sigma(t)$ of the 
cyclic reaction network is expressed in terms 
of fluxes $J_i$ and the corresponding forces $X_i$ as \cite{groot}
\begin{equation}
\sigma(t)=\frac{1}{T}\sum_{i=1}^{3} J_i(t) X_i(t).
\label{epr}
\end{equation}
The fluxes are defined as \cite{groot,nicol,espo}: 
\begin{equation}
J_1(t)=k_1 a(t)-k_{-1}b(t),
\label{j1}
\end{equation}
\begin{equation}
J_2(t)=k_2 b(t)-k_{-2}c(t),
\label{j2}
\end{equation}
\begin{equation}
J_3(t)=k_3 c(t)-k_{-3}a(t).
\label{j3}
\end{equation}
The corresponding forces are 
\begin{equation}
X_1(t)=\mu_A-\mu_B=T{\rm ln}\frac{k_1 a(t)}{k_{-1}b(t)},
\label{x1}
\end{equation}

\begin{equation}
X_2(t)=\mu_B-\mu_C=T{\rm ln}\frac{k_2 b(t)}{k_{-2}c(t)},
\label{x2}
\end{equation}

\begin{equation}
X_3(t)=\mu_C-\mu_A=T{\rm ln}\frac{k_3 c(t)}{k_{-3}a(t)}
\label{x3}
\end{equation}
showing
\begin{equation}
X_1(t)+X_2(t)+X_3(t)=0.
\label{xzero}
\end{equation}
We have set here (and throughout) the Boltzmann constant $k_B=1,$ 
and $T$ refers to the local temperature. 
One can easily see that the reaction velocities 
are related to the fluxes as  
\begin{equation}
\dot{a}=-J_1+J_3,
\label{ratflx1}
\end{equation}
\begin{equation}
\dot{b}=J_1-J_2,
\label{ratflx2}
\end{equation}
\begin{equation}
\dot{c}=J_2-J_3.
\label{ratflx3}
\end{equation}
So, for the cyclic network, {\it none of the reaction velocities 
are equal to the fluxes}. 
At steady state, we also have from Eq.(\ref{ratflx1}-\ref{ratflx3}) 
that 
\begin{equation}
J_1^s=J_2^s=J_3^s=J_c.
\label{fluxss}
\end{equation} 
Then, from Eq.(\ref{epr}) and Eq.(\ref{xzero}), EPR at steady state 
becomes
\begin{equation}
T\sigma=J_c(X_1+X_2+X_3)=0.
\label{eprst}
\end{equation} 
Also, from Eqs (\ref{x1})-(\ref{x3}) and Eq.(\ref{xzero}), we get 
\begin{equation}
\frac{k_1k_2k_3}{k_{-1}k_{-2}k_{-3}}=1.
\label{krelatn}
\end{equation}
The above relation holds when the system satisfies 
the condition of detailed balance\cite{denb,groot}. 
This  
requires the fluxes of each individual 
reaction to vanish at steady state, {\it i.e.}, 
\begin{equation}
J_1^s=J_2^s=J_3^s=J_c=0.
\label{fluxeq}
\end{equation} 
In this case, the reaction system reaches TE. 
Now, using Eq.(\ref{krelatn}), it is easy to verify 
that the steady solutions, Eqs (\ref{as})-(\ref{cs}), do indeed 
satisfy Eq.(\ref{fluxeq}). So the ABC cyclic 
network can only reach TE, and no NESS is possible here.
This is also indicated by the vanishing $\sigma(t)$ given in 
Eq.(\ref{eprst}). 
The TE concentrations are given as
\begin{equation}
a^e=k_{-1}k_3/N_2,
\label{ae}
\end{equation}
\begin{equation}
b^e=k_1k_3/N_2,
\label{be}
\end{equation}
\begin{equation}
c^e=k_{-1}k_{-3}/N_2
\label{ce}
\end{equation} 
where $N_2=k_1k_3+k_{-1}k_3+k_{-1}k_{-3}$.

Consider now a situation when the reaction system is 
close to the TE. 
The concentrations are taken as
\begin{equation}
a(t)=a^e + \delta_a,
\label{adel}
\end{equation}
\begin{equation}
b(t)=b^e + \delta_b,
\label{bdel}
\end{equation}
\begin{equation}
c(t)=c^e + \delta_c
\label{cdel}
\end{equation}
with $\delta_a+\delta_b+\delta_c=0,$ because the sum of 
concentrations of all the species is fixed throughout. 
It is necessary to find out the relations among $\delta_a, \delta_b, 
\delta_c$ to obtain a useful form of EPR 
close to TE. 
From definition, it follows that 
\begin{equation}
\dot{\delta_a}=\dot{a}=-(k_1+k_{-3})\delta_a+k_{-1}\delta_b+k_3\delta_c.
\label{dela}
\end{equation}
From Eq.(\ref{adel}) and Eq.(\ref{dela}), 
one can write for an infinitesimal time interval $\tau$ 
\begin{equation}
(1+(k_1+k_{-3})\tau)\delta_a-k_{-1}\tau\delta_b-k_3\tau\delta_c=0.
\label{deleq1}
\end{equation}
Similarly, from the equations of $\dot{\delta_b}$ and $\dot{\delta_c}$, 
one obtains
\begin{equation}
-k_{1}\tau\delta_a+(1+(k_{-1}+k_{2})\tau)\delta_b-k_{-2}\tau\delta_c=0
\label{deleq2}
\end{equation}
and 
\begin{equation}
-k_{-3}\tau\delta_a-k_2\tau\delta_b+(1+(k_{-2}+k_3)\tau)\delta_c=0.
\label{deleq3}
\end{equation}
Using Eqs.(\ref{deleq1})-(\ref{deleq3}), we get 
\begin{equation}
\delta_b=f_1\delta_a,\quad \delta_c=-(\delta_a+\delta_b)=
f_2\delta_a,
\label{relatn}
\end{equation}
where 
\begin{equation}
f_1=\frac{k_{-2}(1+(k_1+k_{-3})\tau)+k_1k_3\tau}
{k_{-1}k_{-2}\tau+k_3(1+(k_{-1}+k_{2})\tau)};\, f_2=-(1+f_1).
\label{frelatn}
\end{equation}
%and 
%$$f_2=\frac{(1+(k_1+k_{-3})\tau)(1+(k_{-1}+k_{2})\tau)-k_1k_{-1}\tau^2}
%{k_3(1+(k_{-1}+k_{2})\tau)\tau+k_{-1}k_{-2}\tau^2}.$$
It may be pointed out that, in deriving Eq.(\ref{relatn}), we do 
not assume the condition of detailed balance, viz., 
Eq.(\ref{krelatn}).

We next obtain the EPR from Eq.(\ref{epr}) near TE, 
making use of Eq.(\ref{relatn}), the TE concentrations, 
Eqs (\ref{ae}-\ref{ce}) and taking 
$\delta_a,\delta_b,\delta_c$ small, as 
\begin{equation}
\sigma(t)=L_1\delta_a^2,
\label{eprss1}
\end{equation}
where
$$L_1=\left[(k_1-f_1k_{-1})(1/a^e-f_1/b^e)+
(f_1k_2-f_2k_{-2})(f_1/b^e-f_2/c^e)+\right.$$
$$\left.(f_2k_3-k_{-3})(f_2/c^e-1/a^e)\right]$$
$$=\left[\frac{(k_1-f_1k_{-1})^2}{k_1}+\frac{(f_2k_3-k_{-3})^2}{k_{-3}}
\right.$$
\begin{equation}
+\left.\frac{(f_1k_2-f_2k_{-2})(f_1k_{-1}k_{-3}-f_2k_1k_3)}
{k_1k_{-3}}\right]\frac{N_2}{k_{-1}k_3}.
\label{L1}
\end{equation} 
A good cross check at this juncture would be 
to examine whether $L_1$ is positive definite. 
We mention here that the positivity of the last term of Eq.(\ref{L1}), 
and hence the positivity of $\sigma$, is ensured by the 
condition of detailed balance. Indeed, one gets 
from Eq.(\ref{krelatn})
\begin{equation}
\frac{f_1k_{-1}k_{-3}}{f_2k_1k_3}=\frac{f_1k_2}{f_2k_{-2}},
\end{equation} 
and this guarantees the positivity of $L_1$. 

The velocity of the ABC cyclic reaction system, $v(t)$, 
can be expressed as the 
rate of change of concentration of {\it any one} of the three species. 
Let us define $v(t)=\dot{a}.$ Then, close to TE, we have
\begin{equation}
v(t)=(f_2k_3+f_1k_{-1}-(k_1+k_{-3}))\delta_a=L_2\delta_a.
\label{vss}
\end{equation} 
Thus, combining Eq.(\ref{eprss1}) and Eq.(\ref{vss}), we can write
\begin{equation}
\sigma(t)=\frac{L_1}{L_2^2}v^2(t).
\label{eprv}
\end{equation} 
Hence, {\it close to TE, EPR is proportional 
to the square of the rection velocity.} 
It is easy to see that, defining the reaction velocity 
as equal to $\dot{b}$ or $\dot{c}$ generates 
similar type of expression with the same conclusion.  

\subsection{ABC linear network}

\begin{figure}[tbh]
\centering
\rotatebox{270}{
\includegraphics[width=6cm,keepaspectratio]{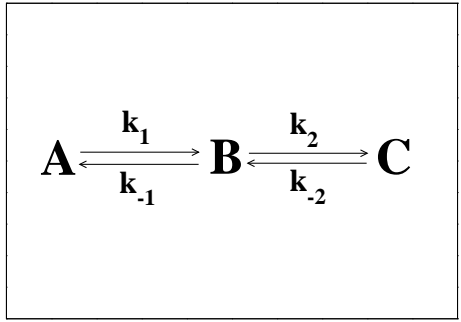}}%{ABC-scheme-lin.ps}}
\caption{Schematic diagram of the ABC linear reaction network indicating 
the forward and backward rate constants of each reaction.}
\label{fig2}
\end{figure}

To emphasize the point expressed in Eq.(\ref{eprv}), we 
take up now the case of the ABC linear reaction network. 
The reaction scheme is given in Fig.\ref{fig2}. 
%%%%%%%%%%%%%%%%%%%%%%%%%%%%%%%%%%%%%%%%%%%%%%%%%%%%%%%%%%%%%%%%%%%%%%
%%%%%%%%%%%%%%%%%%%%%%%%%%%%%%%%%%%%%%%%%%%%%%%%%%%%%%%%%%%%%%%%%%%%%%%%%
The corresponding rate equations and steady state solutions 
can be obtained from Eqs.(\ref{abc1})-(\ref{abc3}) 
by setting $k_3=k_{-3}=0.$ Then, one gets the following 
relations between reaction velocities and fluxes
\begin{equation}
\dot{a}=-J_1,
\label{rateflx1l}
\end{equation}
\begin{equation}
\dot{b}=J_1-J_2,
\label{rateflx2l}
\end{equation}
\begin{equation}
\dot{c}=J_2.
\label{rateflx3l}
\end{equation}
Unlike the case of ABC cyclic network [see Eqs.(\ref{ratflx1})-
(\ref{ratflx3})], here the reaction velocities are 
not all of similar structure. 
Depending on our choice, it can be equal to the flux or 
can be different (see below).

We note first that, the TE solutions of the linear network are 
as follows:
\begin{equation}
a^e=k_{-1}k_{-2}/N_3, 
\label{ael}
\end{equation}
\begin{equation}
b^e=k_{1}k_{-2}/N_3,
\label{bel}
\end{equation}
\begin{equation} 
c^e=k_1k_2/N_3
\label{cel}
\end{equation} 
with $N_3=k_{-1}k_{-2}+k_{1}k_{-2}+k_1k_2.$
The equivalent of Eq.(\ref{relatn}) in this case is
\begin{equation}
\delta_b=f'_1\delta_a,\quad\delta_c=-(\delta_a+\delta_b)=
f'_2\delta_a,
\label{relatnlin}
\end{equation}
with 
\begin{equation}
f'_1=\frac{1+k_1\tau}{k_{-1}\tau},\quad f'_2=-(1+f'_1),
\label{frelatnlin}
\end{equation}
which follows from Eq.(\ref{frelatn}) for $k_3=k_{-3}=0.$
Using the above relations along with $k_3=k_{-3}=0$ in Eq.(\ref{eprss1}), 
the EPR of ABC linear reaction network close to TE becomes
\begin{equation}
\sigma(t)=L_3\delta_a^2,
\label{eprss2}
\end{equation}
where
$$L_3=\left[(k_1-f'_1k_{-1})(1/a^e-f'_1/b^e)+
(f'_1k_2-f'_2k_{-2})(f'_1/b^e-f'_2/c^e)\right]\delta_a^2$$
\begin{equation}
=\left[\frac{(k_1-f'_1k_{-1})^2}{k_{-1}}+
\frac{(f'_1k_2-f'_2k_{-2})^2}{k_2}\right]\frac{N_3}{k_{1}k_{-2}}.
\label{L3}
\end{equation} 
Note that the posistive definite character of $L_3$ is 
transparent. 

Now we define the reaction velocity, say, by $v(t)=\dot{c}$ 
which is equal to the flux $J_2.$ 
Then close to TE, we have 
\begin{equation}
v(t)=(f'_1k_2-f'_2k_{-2})\delta_a=L_4\delta_a.
\label{velin1}
\end{equation}
Therefore, coupling Eq.(\ref{eprss2}) and Eq.(\ref{velin1}), we can write
\begin{equation}
\sigma(t)=\frac{L_3}{L_4^2}v^2(t).
\label{eprvlin1}
\end{equation}
If one chooses to define the velocity as $v(t)=\dot{b},$ which 
is {\it not equal} to any of the fluxes, 
then close to TE one gets
\begin{equation}
v(t)=(k_1+f'_2k_{-2}-f'_1(k_{-1}+k_2))\delta_a=L_5\delta_a.
\label{velin2}
\end{equation}
Consequently, the EPR again becomes 
\begin{equation}
\sigma(t)=\frac{L_3}{L_5^2}v^2(t).
\label{eprvlin2}
\end{equation}
Similar type of quadratic variation follows if one takes $v(t)=\dot{a}.$ 
 
\section{Entropy production rate and reaction rate around 
NESS}

It is now appropriate to take up the cases of chemical reactions 
that {\it can} support a NESS under specified condition. 
This will allow us to investigate whether the relation between EPR 
and reaction velocity near TE, 
derived in Section II, also holds here. 

\subsection{ABC cyclic network}

\begin{figure}[tbh]
\centering
\rotatebox{270}{
\includegraphics[width=8cm,keepaspectratio]{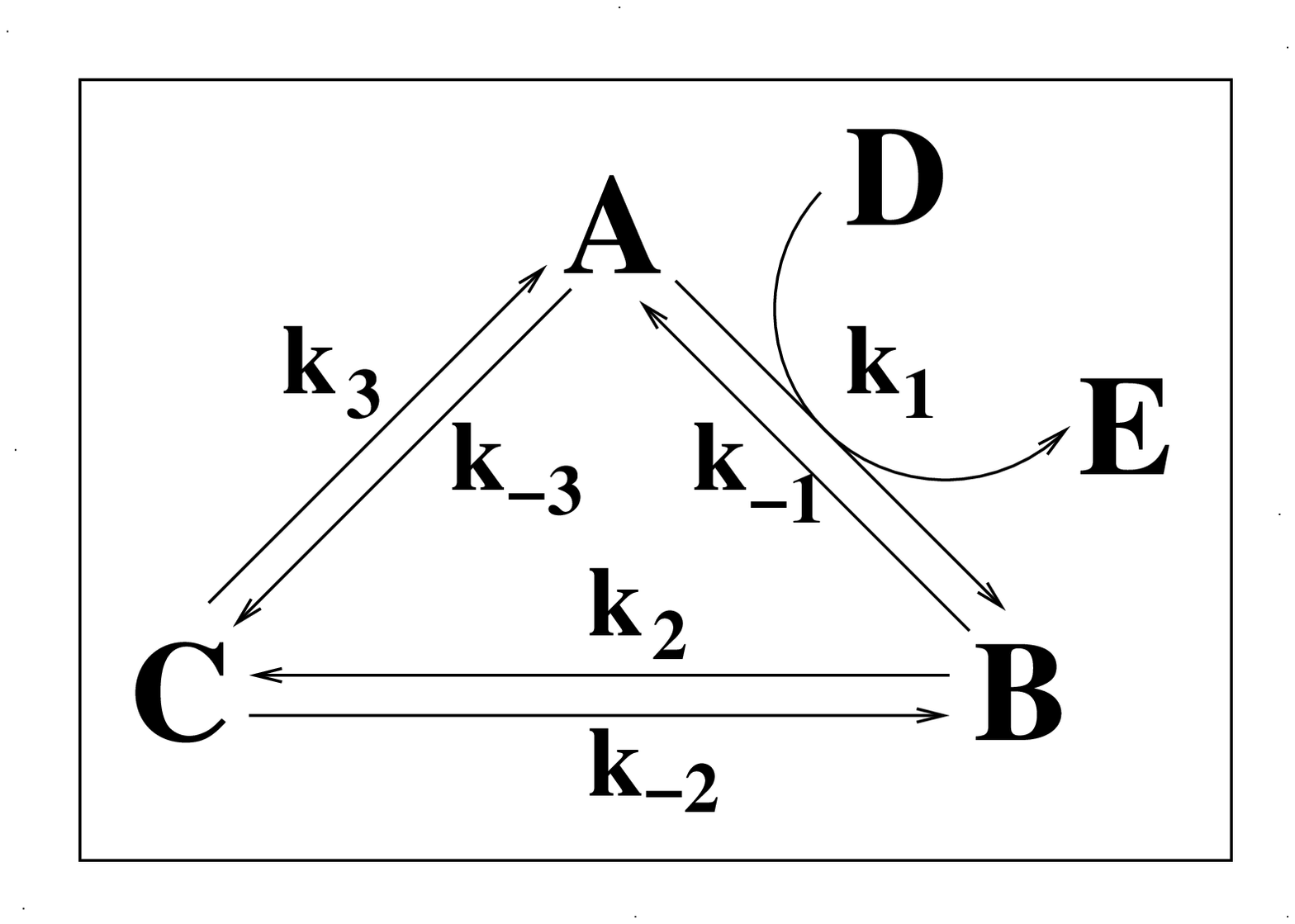}}%{abc-ness.ps}}
\caption{Schematic diagram of the ABC cyclic reaction network 
under chemiostatic condition, with the concentrations of D and E 
held fixed.}
\label{fig3}
\end{figure}

The ABC cyclic reaction network discussed in Section II.A 
does not provide any provision for a NESS.
So, we consider the triangular network 
under a special chemiostatic condition, as shown in Fig.\ref{fig3}. 
Here the concentrations of species D and E are externally 
kept fixed\cite{Hqian2} at $d^0$ and $e^0$. 
%%%%%%%%%%%%%%%%%%%%%%%%%%%%%%%%%%%%%%%%%%%%%%%%%%%%%%%%%%%%%%%%%%%%%%
%%%%%%%%%%%%%%%%%%%%%%%%%%%%%%%%%%%%%%%%%%%%%%%%%%%%%%%%%%%%%%%%%%%%%%%%%
The pseudo-first-order rate constants are defined 
as $k'_1=k_1d^0,$ $k'_{-1}=k_{-1}e^0$. 
Then, the fluxes become 
\begin{equation}
J_1(t)=k'_1 a(t)-k'_{-1}b(t),
\label{j1ne}
\end{equation}
\begin{equation}
J_2(t)=k_2 b(t)-k_{-2}c(t),
\label{j2ne}
\end{equation}
\begin{equation}
J_3(t)=k_3 c(t)-k_{-3}a(t).
\label{j3ne}
\end{equation}
The corresponding forces are 
\begin{equation}
X_1(t)=\mu_A+\mu_D-\mu_B-\mu_E=T{\rm ln}\frac{k'_1 a(t)}{k'_{-1}b(t)},
\label{x1ne}
\end{equation}

\begin{equation}
X_2(t)=\mu_B-\mu_C=T{\rm ln}\frac{k_2 b(t)}{k_{-2}c(t)},
\label{x2ne}
\end{equation}

\begin{equation}
X_3(t)=\mu_C-\mu_A=T{\rm ln}\frac{k_3 c(t)}{k_{-3}a(t)}.
\label{x3ne}
\end{equation}
%with the Boltzmann constant, $k_B$ set equal to 1.
The steady state concentrations will now be 
given still by Eqs.(\ref{as})-(\ref{cs}), 
with $k'_1$, $k'_{-1}$ replacing $k_1$, $k_{-1}$, respectively.
At steady state, the 
fluxes are equal to each other, as was in case of the system 
discussed in Section II.A. 
But, an important difference exists. 
We have here
\begin{equation}
X_1(t)+X_2(t)+X_3(t)=\mu_D-\mu_E=T{\rm ln}
\frac{k'_1k_2k_3}{k'_{-1}k_{-2}k_{-3}}.
\label{forcene}
\end{equation}
Unless the species D and E are in TE, 
the l.h.s. of Eq.(\ref{forcene}) is {\it not} zero. 
This is unlike Eq.(\ref{xzero}) of Section II.A. 
Hence, $\sigma$ will not vanish at the steady state, 
establishing the non-equilibrium nature of the latter with 
broken detailed balance. 
Only when the l.h.s. of Eq.(\ref{forcene}) vanishes, 
we get
\begin{equation}
\frac{k'_1k_2k_3}{k'_{-1}k_{-2}k_{-3}}=1,
\label{dbcst}
\end{equation}
and the NESS  becomes the state of equilibirum satisfying 
detailed balance Eq.(\ref{dbcst}), as appropriate here.

Allowing small deviations in concentration, $\delta_a,\delta_b,
\delta_c$ from the steady state,  
we arrive from Eq.(\ref{epr}) at the general expression of 
$\sigma(t)$ close to the 
NESS in the form
\[\sigma(t)=(J_1^s+k'_1\delta_a-k'_{-1}\delta_b)
\left({\rm ln}\frac{k'_1 a^s}{k'_{-1} b^s}+(\delta_a/a^s-\delta_b/b^s)\right)
\]
\[+(J_2^s+k_2\delta_b-k_{-2}\delta_c)
\left({\rm ln}\frac{k_2 b^s}{k_{-2} c^s}+(\delta_b/b^s-\delta_c/c^s)\right)
\]
\begin{equation}
+(J_3^s+k_3\delta_c-k_{-3}\delta_a)
\left({\rm ln}\frac{k_3 c^s}{k_{-3} a^s}+(\delta_c/c^s-\delta_a/a^s)\right).
\label{eprss}
\end{equation}
We can still use Eq.(\ref{relatn}), now containing the 
pseudo-first-order rate constants, $k'_1$, $k'_{-1}$, 
because its derivation does not 
require the condition of detailed balance. 
Then, using Eq.(\ref{fluxss}) and Eq.(\ref{relatn}), 
we can express Eq.(\ref{eprss}) as 
\begin{equation}
\sigma(t)=P_1+Q_1\delta_a+R_1\delta_a^2,
\end{equation}
where
$$P_1=J_c{\rm ln}\frac{k'_1k_2k_3}{k'_{-1}k_{-2}k_{-3}},$$
$$Q_1=(k'_1-f_1k'_{-1}){\rm ln}\frac{k'_1a^s}{k'_{-1}b^s}+
(f_1k_2-f_2k_{-2}){\rm ln}\frac{k_2b^s}{k_{-2}c^s}+
(f_2k_3-k_{-3}){\rm ln}\frac{k_3c^s}{k_{-3}a^s},$$
$$R_1=L'_1.$$
$L'_1$ has the similar mathematical structure as that of 
$L_1$ in Eq.(\ref{eprss1}) with $a^e,b^e,c^e$ in Eq.(\ref{eprss1}) 
being replaced by $a^s,b^s,c^s$, now containing the 
pseudo-first-order rate constants, $k'_1$, $k'_{-1}$. 

One notes now that the following conditions must hold 
in order that 
the EPR becomes proportional to the square of 
the reaction velocity close to the NESS (see Eq.(\ref{eprv})), 
$$P_1=0=Q_1.$$
However, $P_1=0$ means either 
$$J_c=0$$ and/or 
$$\frac{k'_1k_2k_3}{k'_{-1}k_{-2}k_{-3}}=1.$$ 
Actually these two relations are equivalent, both indicating 
the fulfillment of the detailed balance condition. 
So, when one relation holds, the other becomes automatic. 
Under such a restriction, one finds $Q_1=0$ as $J_c=0.$
So, it follows that $\sigma$ is proportional to the square of 
the reaction 
velocity only near TE, and {\it not} around any NESS, the actual 
relation being already derived in Eq.(\ref{eprss1}).

\subsection{ABC linear network}

The linear ABC network of Section II.B also reaches TE and 
not a NESS. 
This is because, the condition $\dot{a}=\dot{b}=\dot{c}=0$
implies vanishing of all the fluxes at steady state. 
Therefore, it must be a state of TE as 
there is no other option for the system but to obey detailed 
balance. 
Now, if the species A and C are assumed to act 
as chemiostats, {\it i.e.,} their concentrations 
are kept fixed by connecting with external sources, say, 
at values $a^0$ and $c^0$, respectively, then 
a NESS is possible \cite{ross1}. 
The reaction kinetics 
is described by the rate of change of concentration of B as
\begin{equation}
\dot{b}=k_1a^0+k_{-2}c^0-(k_{-1}+k_2)b(t).
\end{equation}
At steady state, $\dot{b}=0$ with 
\begin{equation}
k_1a^0-k_{-1}b^s=k_2b^s-k_{-2}c^0=J_l.
\label{flxcst}
\end{equation}
The NESS solution is then simply
\begin{equation}
b^s=(k_1a^0+k_{-2}c^0)/(k_{-1}+k_2).
\label{bscst}
\end{equation}
However, if we further assume that at steady state, 
\begin{equation}
J_l=0,
\label{dblincst}
\end{equation}
then this corresponds to the condition of detailed balance. The 
system then goes to TE with the concentration 
\begin{equation}
b^e=k_1a^0/k_{-1}=k_{-2}c^0/k_2.
\label{becst}
\end{equation}
This also implies 
\begin{equation}
\frac{k_1k_2a^0}{k_{-1}k_{-2}c^0}=1.
\label{dblcst}
\end{equation}

The expression of EPR is given by
\begin{equation}
\sigma(t)=(k_1a^0-k_{-1}b(t)){\rm ln}\frac{k_1a^0}{k_{-1}b(t)}+
(k_2b(t)-k_{-2}c^0){\rm ln}\frac{k_2b(t)}{k_{-2}c^0}.
\label{eprlinne}
\end{equation}
Now, close to the NESS, with $b(t)=b^s+\delta_b$ as defined earlier, 
it becomes 
\begin{equation}
\sigma(t)=P_2+Q_2\delta_b+R_2\delta_b^2,
\label{eprlincst}
\end{equation}
where
$$P_2=J_l{\rm ln}\frac{k_1k_2a^0}{k_{-1}k_{-2}c^0},$$
$$Q_2=\left(k_2{\rm ln}\frac{k_2b^s}{k_{-2}c^0}-
k_{-1}{\rm ln}\frac{k_1a^0}{k_{-1}b^s}\right),$$
$$R_2=(k_{-1}+k_2)/b^s.$$ 
On the other hand, the reaction velocity, $v(t)=\dot{b}$ close to the 
steady state becomes 
\begin{equation}
v(t)=-(k_{-1}+k_2)\delta_b.
\label{vlincst}
\end{equation}

Therefore, for $\sigma$ to be proportional to the square of the 
reaction velocity, one needs 
$$P_2=0=Q_2.$$ 
Setting $P_2=0$ means either $$J_l=0$$ 
and/or 
$$\frac{k_1k_2a^0}{k_{-1}k_{-2}c^0}=1.$$ 
However, as shown above, the first condition implies the 
second one, and the system satisfies detailed balance. 
With $J_l=0$, we also find that $Q_2=0.$
Hence, it is verified that the proportionality between EPR 
and reaction velocity squared is valid when the reaction 
system is near TE and {\it not} a NESS. 
The final expression of $\sigma$ in the former case becomes  
\begin{equation}
\sigma(t)=(k_{-1}+k_2)\frac{\delta_b^2}{b^e}=
\frac{v^2(t)}{b^e(k_{-1}+k_2)}.
\label{eprlincst1}
\end{equation}

To summarize the results obtained so far, the EPR 
is shown to be proportional to the square of the reaction 
velocity only near TE and not any arbitrary 
NESS. This feature can act as a measure to distinguish between 
a TE and a NESS.

\section{Link with the minimum entropy production principle}

Before concluding, we investigate any possible 
connection between the behavior of EPR 
near a NESS and the MEPP. 
The reasons behind such an endeavour are twofold. 
The first point is that, recently 
it has been shown rigorously by Ross and coauthors \cite{ross1}, 
taking heat flow and chemical reactions as 
examples of non-equilibrium processes, 
that MEPP is true if and only if a steady state is the state of TE 
\cite{ross2}. 
So MEPP can theoretically distinguish a NESS from a TE. 
The second point arises because, the mathematical expressions 
of EPR in the various cases considered in Section II 
and Section III are derived by expanding it around TE and a NESS, 
respectively. 
Such a type of expansion is also used to find 
the extremum of the quantity at that point. For non-negative 
EPR, this extremum is obviously the 
minimum. 
Therefore, here we investigate the validity of MEPP using the 
expressions of EPR 
in cyclic and linear networks reaching NESS under 
chemiostatic condition, as discussed in Section III.

First we take the ABC cyclic reaction network 
under chemiostatic condition, discussed in Section III.A. 
From the definitions of fluxes and forces 
(Eqs (\ref{j1ne})-(\ref{x3ne})), we find at NESS
\begin{equation}
\left(\frac{\partial\sigma}{\partial a}\right)_s=
k'_1{\rm ln}\frac{k'_1a^s}{k'_{-1}b^s}-k_{-3}{\rm ln}
\frac{k_3c^s}{k_{-3}a^s},
\label{eprmin1}
\end{equation}

\begin{equation}
\left(\frac{\partial\sigma}{\partial b}\right)_s=
k_2{\rm ln}\frac{k_2b^s}{k_{-2}c^s}-k'_{-1}{\rm ln}
\frac{k'_1a^s}{k'_{-1}b^s},
\label{eprmin2}
\end{equation}

\begin{equation}
\left(\frac{\partial\sigma}{\partial c}\right)_s=
k_3{\rm ln}\frac{k_3c^s}{k_{-3}a^s}-k_{-2}{\rm ln}
\frac{k_2b^s}{k_{-2}c^s}.
\label{eprmin3}
\end{equation}
Now extremum of $\sigma$ at NESS [which is obviously the 
minimum, as $\sigma\ge 0,$] requires 
\begin{equation}
\left(\frac{\partial\sigma}{\partial a}\right)_s=
\left(\frac{\partial\sigma}{\partial b}\right)_s=
\left(\frac{\partial\sigma}{\partial c}\right)_s=0.
\label{mincon}
\end{equation}
Then from Eq.(\ref{eprmin1}) and Eq.(\ref{eprmin2}), 
we get the condition
$$\frac{k'_1a^s}{k'_{-1}b^s}=\left(\frac{k_3c^s}{k_{-3}a^s}\right)
^{k_{-3}/k'_1}=
\left(\frac{k_2b^s}{k_{-2}c^s}\right)^{k_2/k'_{-1}}.$$
Hence
\begin{equation}
\frac{k_3c^s}{k_{-3}a^s}=
\left(\frac{k_2b^s}{k_{-2}c^s}\right)^{k'_1k_2/k'_{-1}k_{-3}}.
\label{con1}
\end{equation}
From Eq.(\ref{eprmin3}), one further gets
\begin{equation}
\frac{k_3c^s}{k_{-3}a^s}=
\left(\frac{k_2b^s}{k_{-2}c^s}\right)^{k_{-2}/k_3}.
\label{con2}
\end{equation}
Comparing the right hand sides of Eq.(\ref{con1}) and 
Eq.(\ref{con2}), we get
$$\frac{k'_1k_2k_3}{k'_{-1}k_{-2}k_{-3}}=1.$$
The above condition is fulfilled when the ABC cyclic reaction network 
obeys detailed balance. Therefore, it is seen that 
the NESS must be the state of TE to have minimum EPR, 
as emphasized by Ross {\it et. al.}\cite{ross1}. 

Now coming to the ABC linear network, discussed in Section III.B., 
we obtain from Eq.(\ref{eprlinne}) 
\begin{equation}
\left(\frac{\partial\sigma}{\partial b}\right)_s=
k_2{\rm ln}\frac{k_2b^s}{k_{-2}c^0}-
k_{-1}{\rm ln}\frac{k_1a^0}{k_{-1}b^s},
\label{eprminlin}
\end{equation}
at NESS.
Setting $\left(\frac{\partial\sigma}{\partial b}\right)_s=0$, 
we get
\begin{equation}
b^s=\left(\frac{k_1a^0}{k_{-1}}\right)^{\frac{k_{-1}}{k_{-1}+k_2}}
\left(\frac{k_{-2}c^0}{k_2}\right)^{\frac{k_2}{k_{-1}+k_2}}.
\label{bsmin}
\end{equation}
Now putting the expression of $b^e$ from Eq.(\ref{becst}) in 
Eq.(\ref{bsmin}), one finds
\begin{equation}
b^s=\left(b^e\right)^{\frac{k_{-1}}{k_{-1}+k_2}}
\left(b^e\right)^{\frac{k_2}{k_{-1}+k_2}}=b^e.
\end{equation}
Thus, the EPR is again a minimum only at TE.

\section{Conclusion}

Focusing particularly on 
chemical reactions, in this endeavor, 
we have established a connection between 
the EPR and chemical reaction rate. 
Both cyclic and linear networks are considered that 
can attain either a TE or a NESS. 
We have shown that the EPR 
in these systems is proportional to the square 
of the reaction velocity around TE. 
We have further established that the result is {\it not} valid around 
a NESS. 
Hence, our result can be used to theoretically differentiate 
a NESS from a TE. 
Another way is provided by Ross {\it et al.} \cite{ross1} 
that relies on the behavior of the MEPP. 
Thus, the two features, viz., 
(i) proportionality of EPR to the square of the 
reaction velocity near a TE 
and 
(ii) EPR having its minimum at that TE, 
have a common thread. 
Both of them are invalid when the state is a NESS. 
Our findings should be generalizable to more complex reaction 
networks and such studies will be reported in due course.

\section*{Acknowledgment}

K. Banerjee acknowledges the University Grants Commission (UGC), India 
for Dr. D. S. Kothari Fellowship. The authors 
are grateful to Prof. D. S. Ray for a thorough discussion.

\newpage

\section*{Notes}

%\noindent
1. The EPR $\sigma(t)$ in Eq.(\ref{epr}) should be written, 
more precisely, as
$$\sigma(t)=\frac{1}{T}\sum_{i=1}^{3} J_i^{\rm kin}(t) X_i(t),
\qquad \qquad \qquad \qquad \quad(7')$$
$$\sigma^{\rm th}(t)=\frac{1}{T}\sum_{i=1}^{3} J_i^{\rm kin}(t) 
X_i^{\rm th}(t),\qquad \qquad \qquad \qquad (7a)$$
$$\sigma^{\rm kin}(t)=\frac{1}{T}\sum_{i=1}^{3} J_i^{\rm kin}(t) 
X_i^{\rm kin}(t).
\qquad \qquad \qquad \qquad (7b)$$
$J_i$ is always kinetic in nature, as in Eqs (\ref{j1})-(\ref{j3}). 
Eq.(\ref{x1}) is more precisely 
$$X_1^{\rm th}(t)=\mu_A-\mu_B,\qquad \qquad \qquad \qquad \qquad
\qquad(11a)$$
$$X_1^{\rm kin}(t)=T{\rm ln}\frac{k_1 a(t)}{k_{-1}b(t)}.
\qquad \qquad \qquad \qquad \qquad \quad(11b)$$
Similar definitions apply to Eqs (\ref{x2})-(\ref{x3}). 

The distinctions may be appreciated in view of the folllowing:

\noindent
(a) $J_i$ cannot be written as a linear combination of 
$X_i^{\rm th}$ near equilibrium. Thus, Onsager linear relations are  recovered only when one uses $X_i^{\rm kin}$. 

\noindent
(b) The equality appearing in Eqs (\ref{x1})-(\ref{x3}) 
rests on the additional assumption of van't Hoff:  
An equilibrium constant between a reactant and a product is expressible 
as a ratio of forward and backward rate constants. 

\noindent
(c) It is a standard convention to define a steady state (SS) 
as a state where all time-dependences in observables vanish. 
Hence, here, fluxes are equal, as in Eq.(\ref{fluxss}). 
A TE state, on the other hand, is the one with all forces, 
$X_i^{\rm th}$, equal to zero.

\noindent
(d) Note, however, that $\sigma^{\rm th}({\rm SS}) = 0$ 
because $\sum_i X_i^{\rm th} = 0$ (cf. Eq.(\ref{xzero})) 
and all fluxes are equal. 
But, $\sigma^{\rm kin}({\rm SS})$ is {\it not} equal to zero. 
Its vanishing at SS will be ensured only when detailed balance (DB) 
is obeyed.

\noindent
(e) In view of (d), one observes that the role of DB is important only when (b) is assumed {\it a priori}.

%\noindent
2. Reaction rate plays a premier role in the present endeavor. Its link with the EPR that we have established is specific to chemical reaction systems. Such a kinship is difficult to obtain in a general way because the thermodynamic forces may not always be easily expressible in terms of the kinetic ones, as has been accomplished here in Eqs (\ref{x1})-(\ref{x3}).

%\noindent
3. In going from Eq.(\ref{dela}) to Eq.(\ref{deleq1}), 
we have invoked the finite difference approximation to the differential.  Thus, $\dot{\delta_a}=\dot{a}\approx\delta_a/\tau.$

%\noindent
4. The quantity $\tau$ in Eq.(\ref{deleq1}) refers to a time before the  attainment of a SS or a TE. 
Hence, the above association is not in anyway connected to a Taylor expansion. We choose the SS (or TE) at $t = 0$ and consider a 
time $\tau$ before it (i.e., $t = - \tau$), so that one is close to SS, 
but not exactly at it. A Taylor expansion around SS is not permissible  because no change in observables at any $t > 0$ is allowed. 

%\noindent
5. We have actually two independent variables in the cyclic 
triangular reaction system. But, it will be unwise to conclude 
on the basis of Eq.(\ref{relatn}) that $\delta_b$ and 
$\delta_a$ are dependent. Indeed, they are independent. 
The connection via $f_1$ shows only that $\delta_b$ cannot be 
{\it arbitrary} for some given $\delta_a$. Note that $f_1$ 
contains the characteristic reaction constants plus the 
time gap $\tau$. At a different $\tau$, $f_1$ will change, 
thus altering $\delta_b$, even if $\delta_a$ is held fixed. 

As an example, consider the triangular system with all 
rate constants ($k_1$ to $k_{-3}$) equal to unity. 
The conventional solutions (initial condition at $t=0$) 
read as 
$$a(t) = (1/3)-(1/3)(1-3a_0)\,{\rm exp}\,[-3t]$$
$$b(t) = (1/3)-(1/3)(1-3b_0)\,{\rm exp}\,[-3t].$$
The variables $a(t)$ and $b(t)$ {\it are} independent. 
However, after a time $\tau$, one will find that 
$$a(\tau) = a_0+\tau(1-3a_0)$$
$$b(\tau) = b_0+\tau(1-3b_0).$$
Therefore, one can write that 
$$b(\tau) = f_1a(\tau)$$
with $$f_1=(b_0+\tau(1-3b_0))/(a_0+\tau(1-3a_0)).$$
We thus see that $f_1$ merely links the changes 
of two independent variables. It should not be confused 
with a {\it proportionality constant}.  

6. The present work is not an application of the MEPP. 
Rather, it provides an alternative characterization of NESS 
vs. TE. The MEPP also distinguishes these two kinds 
of states. So we explored any possible connection of our 
endeavor with the MEPP.

\end{document}